\documentclass[aps,prb,twocolumn,groupedaddress]{revtex4}
\usepackage{graphicx}
\usepackage{epstopdf}
\usepackage{amsmath}
\begin{document}

\title{Phononic crystal waveguides for electromechanical circuits}

\author{D. Hatanaka}

\email{hatanaka.daiki@lab.ntt.co.jp}

\author{I. Mahboob}

\author{K. Onomitsu}

\author{H. Yamaguchi}

\affiliation{NTT Basic Research Laboratories, NTT Corporation, Atsugi-shi, Kanagawa 243-0198, Japan}

%\begin{abstract}
%\end{abstract}

\maketitle
\hspace*{0em}\textbf{Nanoelectromechanical systems (NEMS), utilising localised mechanical vibrations, were pioneered for sensors\cite{chaste_sensor, arlett_sensor}, signal processors\cite{loh_nems, imran_memory} and to study macroscopic quantum mechanics\cite{connell_qgs, teufel_qgs, chan_qgs}. Increasingly the concept of integrating multiple mechanical elements via electrical/optical means has emerged as a challenge towards NEMS circuits\cite{heinrich_array, stannigel_array, zhang_array, massel_array}. Here we develop phononic crystal waveguides (PnC WGs), using a 1-dimensional array of suspended membranes, that offer purely mechanical means to integrate isolated NEMS resonators. In the first steps to this objective, we demonstrate the PnC WG can support and guide mechanical vibrations where the periodic membrane arrangement also creates a phonon bandgap that enables the phonon propagation velocity to be controlled. Additionally embedding a phonon-cavity into the PnC allows mechanical vibrations in the WG to be dynamically switched or transferred to the cavity, illustrating the availability of WG-resonator coupling. These highly functional traits of the PnC WG architecture exhibit all the ingredients necessary to permit the realisation of all-phononic NEMS circuits.}\\
\hspace*{0.3cm}NEMS resonator circuits can provide the means to synchronise isolated mechanical resonators leading to highly precise oscillators\cite{stannigel_array, zhang_array} as well as suggesting a framework that can couple multiple quantum systems for quantum technology applications\cite{stannigel_array, massel_array}. An intuitive approach to NEMS circuits is to utilise mobile mechanical vibrations, for example in phonon crystal waveguides, to couple localised NEMSs.\\
\hspace*{1em}Phononic crystals are the acoustic analogue of photonic crystals and they herald the possibility of confining and guiding phonons. However, in spite of this promise little effort has been devoted to developing artificial media designed to control phonon propagation. Although passive PnCs have recently been demonstrated\cite{benchabane_pnc, laude_pnc, sun_pnc, mohammadi_pnc2}, dynamic control of phonons which is a necessary ingredient for NEMS circuits has remained elusive. Central to the success of photonic crystals is their ability to dynamically control photons by either the thermo-optic\cite{vlasov_ptc} or the carrier-plasma effects\cite{sato_ptc, tanabe_vol_ptc, nozaki_switch_ptc}. In this report, a dynamic phonon-based analogue is realised in a PnC WG.\\
\hspace*{1em}The PnC WG is composed of a 1-dimensional array of mechanically-coupled membranes as shown in Fig. 1a. The membranes are made from GaAs/AlGaAs heterostructure that are suspended above the GaAs substrate as described in Methods. An electromechanical transducer is formed between the ${\it n}$-type GaAs layer and the gold electrode on the edge membranes which can electrically excite mechanical oscillations from the piezoelectric effect\cite{hatanaka_memb, hatanaka_memarray}.\\
\hspace*{1em}First the transmission of phonons namely mechanical oscillations in the PnC WG are investigated. In these experiments, mechanical oscillations are piezoelectrically excited from the right edge membrane and the propagation of the resultant mechanical oscillations along the array is detected on the left edge membrane via optical interferometry. The results of these measurements are shown in Fig. 1b for a range of PnC WGs whose constituent membrane number (${\it N}$) is varied from 10, 20, 50 and 100. The transmission spectrum of the PnC WG with ${\it N}$ = 10 exhibits equally spaced peaks which result from Fabry-Perot resonances arising from the boundaries of the WG at the left and right edge membranes. However as ${\it N}$ increases, the peak separation reduces and eventually forms phonon bands (${\it N}$ $\geq$ 50). Moreover, a phonon bandgap also emerges in this regime due to the half wavelength of the phonon standing waves matching the membrane separation\cite{kittel}.\\
\hspace*{1em}To further evaluate the properties of the PnC WGs, their temporal dynamics are investigated. The propagation of phonons, excited at the right edge of the PnC WGs with a pulse width of 1 $\mu$s for a range of frequencies, is measured at the left edge of the WGs as a function of time and is shown in Figs. 2a-2d. These measurements reveal mechanical oscillations originating at the right edge membrane propagate through the PnC WGs and are reflected at the left edge membrane. Indeed, multiple reflections are observed as evidenced by the phonon propagation fringes in Figs. 2a-2d. The resultant fringe period corresponds to a round trip and its duration is determined by the length of the PnC WG, where clearly a larger period is observed when ${\it N}$ increases. Additionally, the phonon transmission in the PnC WG is confined to the spectral regions in which the bands exist and is suppressed in the bandgap as shown in Fig. 2c and Fig. 2d. Moreover the presence of higher-order dispersion also broadens the phonon pulse as it travels down the PnC WG and an example of this is shown in Fig. 2e and is detailed in section S2 in Supplementary Information (SI).\\
\hspace*{1em}In order to understand both the spatial and temporal characteristics of the phonon pulse propagation, the probe laser is swept along the PnC WG with ${\it N}$ = 100 at 5.5 MHz as shown in Fig. 2f. This measurement reveals that phonons propagate with a constant speed with reflection taking place at the WG's edges. Analysis of the phonon pulse propagation track yields a group velocity (${\it v}$$_{\rm g}$) of 125 m/s with an energy loss of 3 dB/mm. These results indicate that phonons can be efficiently confined and guided by the PnC WG.\\
\hspace*{1em}The speed at which the phonons travel through the PnC WG is determined by spectral structure of the phonon bands. The corresponding group velocities can be determined from both the spectral and temporal measurements shown in Fig. 1b and Figs. 2a-2d respectively\cite{baba_ptc}. Specifically, in the spectral measurements, the peak separation $\Delta$${\it f}$ between the Fabry-Perot resonances is defined by $\Delta$${\it f}$ = ${\it v}$$_{\rm g}$/2${\it L}$ (where ${\it L}$ is the length of the WG). On the other hand, the fringe separation $\Delta$${\it t}$ in the temporal measurements is determined by $\Delta$${\it t}$ = 2${\it L}$/${\it v}$$_{\rm g}$. The extracted group velocities clearly exhibit a frequency dependence as shown in Fig. 3 and in the proximity of the bandgap are reduced by a factor of 2. This observation of ${\it slow}$ ${\it phonons}$\cite{laude_pnc, sun_pnc} is analogous to ${\it slow}$ ${\it photons}$ and is a consequence of the large change in the phonon dispersion relation near the band edges\cite{baba_ptc}.\\
\hspace*{1em}In order to confirm this assertion, the experimental results in Fig. 3 are reproduced by using a model for the PnC WG composed of an array of coupled articulated beams\cite{watanabe_wavemotion} as described in section S1 in SI. Specifically, this model enables the application of the Euler-Bernoulli formalism to describe the dynamics of the beams and, it permits the corresponding dispersion relation for the PnC WG to be determined. The group velocities can then be extracted from this dispersion relation as outlined in section S1 in SI. The experimental and calculated group velocities converge when ${\it N}$ increases, as shown in Fig. 3. Fundamentally, this analysis verifies that a bandgap emerges for a sufficiently large ${\it N}$ in the PnC WG and the resultant reduction in the group velocity is the natural consequence of the rapidly varying dispersion relation in the proximity of the bandgap.\\
\hspace*{1em}Although the phonon transmission can be controlled by exploiting the dispersion relation in the proximity of the bandgap, the PnC WG is fundamentally a passive system. To realize the full potential of PnC WGs in an active configuration, a cavity is embedded into the centre of the WG by varying the spacing of the membranes with ${\it N}$ = 100 as shown in Fig. 1c. This partially isolated membrane sustains a local vibration mode around 2 MHz which plays the role of a cavity as shown in the red arrow in Fig. 1d. The phonon transmission in the PnC WG with cavity is measured as before and is shown in Fig. 1e. This measurement reproduces the spectral response of the PnC WG with ${\it N}$ =100 except now the phonon transmission between 2-3 MHz is suppressed. The introduction of the cavity results in an acoustic impedance mismatch with the WG resulting in phonons  being reflected at the WG/cavity interface (see section S3 in SI). Consequently, by varying the dimensions thus the frequency of the cavity membrane, the PnC WG transmission spectrum can be controllably engineered. However, the cavity can also enable dynamic control of the phonon transmission in the PnC WG when it is directly excited.\\
\hspace*{1em}To demonstrate such active control, the PnC WG with cavity is excited at 5.745 MHz (see orange arrow in Fig. 1e) while simultaneously, the cavity is excited around 1.87 MHz (see red arrow in Fig. 1d). Excitation of the cavity mechanically modulates the spring constant of the PnC WG resulting in its frequency blue shifting\cite{karabalin_nonlinear, westra_nonlinear, gaidarzhy_nonlinear}. Indeed, when the cavity is excited on resonance, the PnC WG mode at 5.745 MHz no longer exists as shown in Fig. 4a. Consequently, the transmission in the PnC WG can be dynamically switched by pulsing the cavity as shown in Fig. 4b.\\
\hspace*{1em}On the other hand, the off-resonance modulation of the cavity membrane can enable transfer of vibrations from the WG to the cavity. As before the PnC WG is excited at 5.745 MHz but now the cavity is simultaneously excited far from its resonance at 3.809 MHz (see green arrow in Fig. 1d), namely at the frequency difference to the WG mode. Excitation of the cavity at this frequency difference can modulate the tension of the PnC WG and it results in the generation of a sideband close to the WG mode at 5.745 MHz\cite{hatanaka_memarray, imran_eit, teufel_eit}. The interaction of the cavity-induced sideband and the PnC WG mode results in energy being transferred from the PnC WG to the cavity as evidenced by the emergence of a dip in the response of the WG mode as shown in Fig. 4c (see also section S4 in SI). This phenomenon can also be exploited to impart temporal control to the PnC WG as shown in Fig. 4d. In contrast to the dispersion of the PnC WG mode by the cavity as described above (see Fig. 4a and 4b), this effect enables the PnC WG to couple to the cavity, i.e. a localised NEMS resonator.\\
\hspace*{1em}The PnC WG heralds artificially-engineered periodic elastic structures that can both confine and guide phonons. Consequently, these demonstrations strongly suggest that concepts from photonic crystals can be readily applied to phonons. Most promisingly, it is anticipated that PnC WGs could spatially transfer mechanical vibrations between multiple NEMS resonators with complete dynamic control thus opening up the prospect of all-phononic NEMS circuits.\\
\\
\noindent{\textbf{Methods}}\\
\hspace*{1em}{\small The devices were fabricated from a GaAs (5 nm) / Al$_{\rm 0.27}$Ga$_{\rm 0.73}$As (95 nm) / ${\it n}$-type GaAs (100 nm) / Al$_{\rm 0.65}$Ga$_{\rm 0.35}$As (3.0 $\mu$m) heterostructure on GaAs substrate using conventional micromachining processes. The membranes were suspended by removing the Al$_{\rm 0.65}$Ga$_{\rm 0.35}$As sacrificial layer through the periodic air-hole array with a 5\% hydrofluoric acid solution buffered with water for 33 minutes$^{20}$.\\
\hspace*{1em}The mechanical oscillations in the PnCs are excited by applying an alternating voltage from a signal generator (NF Wavefactory 1974) which are then optically detected by a He-Ne laser Doppler interferometer (Neoark MLD-230V-200). The electrical output from the interferometer is then measured with a vector signal analyzer (HP 89410A), a lock-in amplifier (SRS SR844) or an oscilloscope (Agilent DSO6014A).\\
\hspace*{1em}In the temporal measurements in Fig. 4b and 4d, the output from the interferometer is first filtered with a lock-in amplifier (Zurich instruments HF2LI) followed by a low pass filter with a cut-off frequency of 10 kHz (NF 3627) and is then measured in the oscilloscope. This measurement configuration enables a specific vibration in the PnC WG to be selected and then temporally analysed. The extrinsic time delays from both the lock-in amplifier and the low-pass filter have been subtracted from the data in Fig. 4b and 4d.}\\
%
%\bibliography{ref_pnc}

\vspace{0.2cm}
\noindent\textbf{{\small Acknowledgements}} {\small The authors are grateful to Y. Ishikawa for growing the heterostructure and T. Watanabe for helping with the sample preparation. This work was partially supported by JSPS KAKENHI (23241046).}\\
\\
\textbf{{\small Author contributions}} {\small D.H. conceived the experiment, fabricated the  sample, performed measurements and data analysis. K.O. co-fabricated the GaAs/AlGaAs heterostructure. D.H. and I.M. wrote the paper based on discussions with H.Y. who also planned the project.}\\
\\
\textbf{{\small Additional information}} {\small The authors declare no competing financial  interests. Supplementary information accompanies this paper at www.nature.com/naturenanotechnology. Reprints and permissions information is available at www.nature.com/reprints. Correspondence and requests for materials should be addressed to D.H.}

\newpage

\begin{figure*}[floatfix]
\begin{center}
\renewcommand{\figurename}{Figure}
\vspace{3cm}\hspace{8cm}
\includegraphics[scale=0.8]{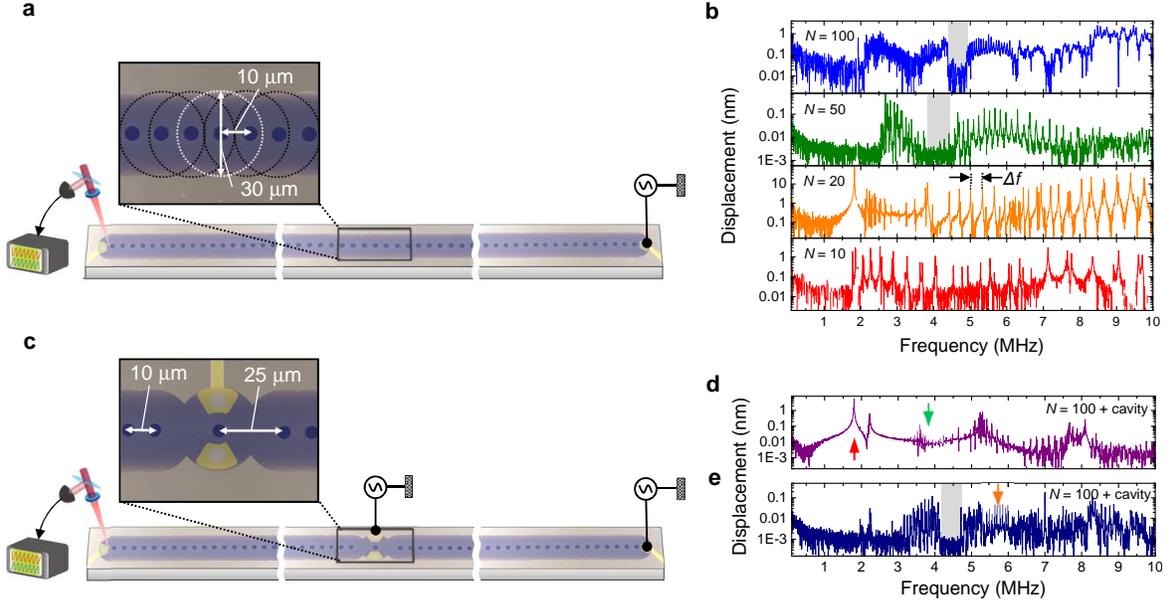}
\vspace{-15cm}
\caption{\textbf{Mechanical membrane-based phononic crystal waveguides.} ${\bf a}$ and ${\bf c}$, Electron micrographs of the PnC WG without and with a cavity (see insets) respectively and the corresponding measurement set-up. The PnC WGs are formed by an array of circular membranes with a diameter of 30 $\mu$m (see inset ${\bf a}$) which are suspended via the circular aperture and are strongly coupled to their neighbouring membranes via the small spatial separation of only 10 $\mu$m. The cavity is embedded into the PnC WG by simply increasing the separation of one of the membrane to its neighbours to 25 $\mu$m as shown in the inset to $\bf{c}$. In all cases, the PnC WGs are piezoelectrically excited at the right edge and the resultant phonon propagation is detected via laser Doppler interferometry at room temperature and in high vacuum. In the PnC WG with cavity, the cavity membrane can also be locally excited via the piezoelectric effect. ${\bf b}$, The frequency response of the PnC WGs with ${\it N}$ = 10, 20, 50 and 100 piezoelectrically excited with an amplitude of 0.4, 1.0, 1.0 and 0.4 V$_{\rm rms}$, respectively and probed at the left most membrane. ${\bf d}$, The frequency response of the cavity in the PnC WG with ${\it N}$ = 100 when locally excited and probed with an amplitude of 1.0 V$_{\rm rms}$. ${\bf e}$, The frequency response of the PnC WG with cavity when excited at the right most membrane with an amplitude of 1.0 V$_{\rm rms}$ and probed at the left most membrane.}
\label{fig 1}
\vspace{-0.7cm}
\end{center}
\end{figure*}

\begin{figure*}[floatfix]
\begin{center}
\renewcommand{\figurename}{Figure}
\vspace{3cm}\hspace{8cm}
\includegraphics[scale=0.8]{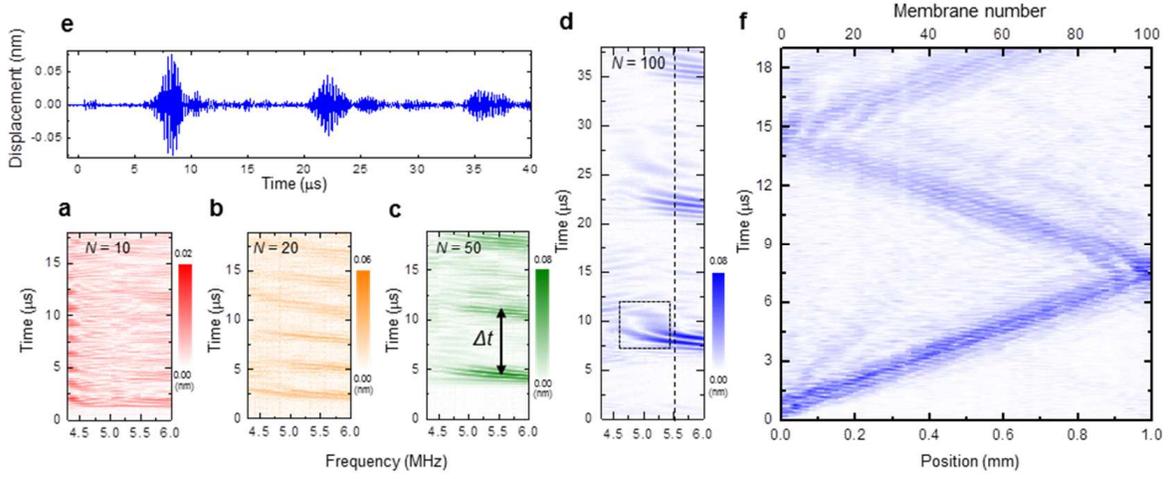}
\vspace{-16.3cm}
\caption{\textbf{Temporal and spatial dynamics of the phonon propagation.} ${\bf a}$-${\bf d}$, The PnC WGs with ${\it N}$ = 10, 20, 50 and 100 excited at the right edge membrane with an amplitude and pulse width of 1.0 V$_{\rm rms}$ and 1 $\mu$s results in mechanical oscillations that are detected at the left edge membrane as a function of time and excitation frequency. As the mechanical oscillations propagate through the PnC WGs, they are detected at the left edge which results in a fringe whose period increases as ${\it N}$ increases. For excitation frequencies corresponding to the bandgap (${\it N}$ $\geq$ 50), the mechanical oscillations thus the phonon propagation fringes are suppressed. ${\bf e}$, The mechanical oscillation detected at the left edge membrane in the PnC WG with ${\it N}$ = 100 at 5.5 MHz from a 1 $\mu$s input pulse (dashed line in ${\bf d}$). ${\bf f}$, The position dependence of the same mechanical oscillations as the probe laser is swept along the membrane array.}
\label{fig 2}
\vspace{-0.7cm}
\end{center}
\end{figure*}

\begin{figure*}[floatfix]
\begin{center}
\renewcommand{\figurename}{Figure}
\vspace{3cm}\hspace{8cm}
\includegraphics[scale=0.5]{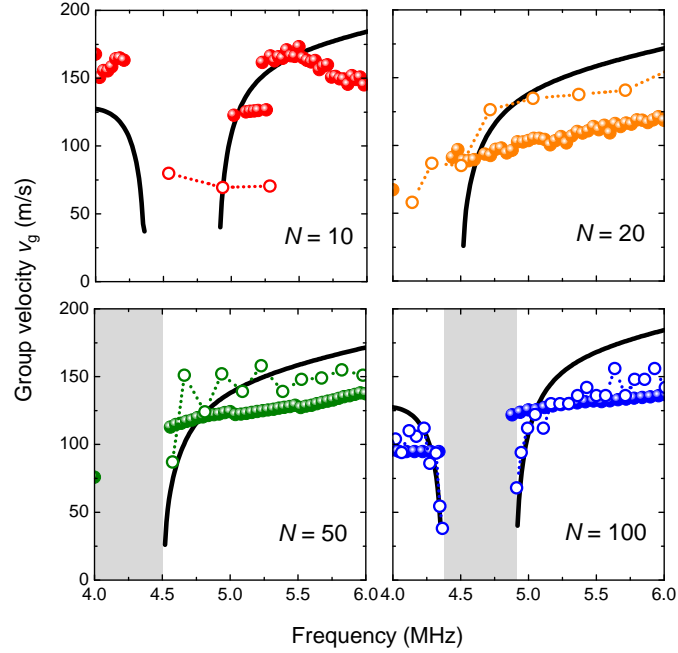}
\vspace{-4.5cm}
\caption{\textbf{The phonon group velocities.} The frequency response of the phonon group velocities in the PnC WGs are determined from the Fabry-Perot resonances (open circles) and from the temporal measurements (spheres) respectively. The bold lines are extracted from the model described in section S1 in SI. The spectral measurements are confirmed by the model calculation where suppression of the group velocity in the proximity of the band edges is reproduced when ${\it N}$ $\geq$ 50. On the other hand, the temporal measurements do not exhibit a clear reduction in the group velocity around the band edges. This is due to the short 1 $\mu$s pulse which results in a spectrally broad wave packet that gives rise to higher-order dispersion in the PnC WG. Consequently, the temporal measurements become distorted exhibiting interference in the proximity of the band edges as highlighted by the box in Fig. 1d.}
\label{fig 3}
\vspace{-1.0cm}
\end{center}
\end{figure*}

\begin{figure*}[floatfix]
\begin{center}
\renewcommand{\figurename}{Figure}
\vspace{3cm}\hspace{0cm}
\includegraphics[scale=0.5]{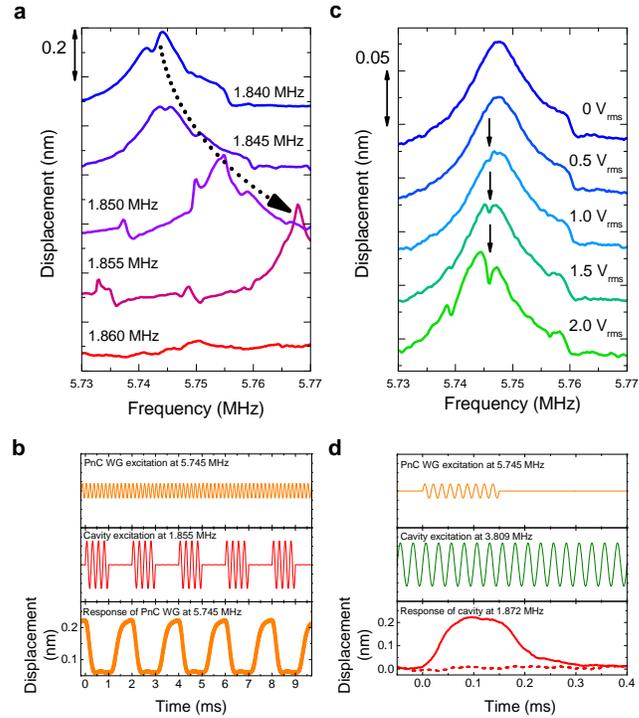}
\vspace{-4.5cm}
\caption{\textbf{Dynamic control of phonons in the PnC WG via the cavity.} ${\bf a}$, The response of the PnC WG excited with an amplitude of 0.4 V$_{\rm rms}$ from the right membrane and probed at the left membrane. Simultaneously, the cavity membrane is excited at 2.0 V$_{\rm rms}$ and as its frequency is swept around its resonance it induces a blue shift of the PnC WG mode. ${\bf b}$, This effect can be used to build a switch by periodically exciting the cavity on resonance at 1.855 MHz whilst the PnC WG mode at 5.745 MHz is continuously excited. The resultant response in the lower panel clearly shows phonon switching in the PnC WG. ${\bf c}$, The frequency response of the PnC WG mode when the cavity is excited at 3.809 MHz whilst its amplitude is increased from 0 to 2.0 V$_{\rm rms}$. This results in the emergence of a dip in the response of the PnC WG mode (see arrows) due to its mechanical vibration energy being transferred to the cavity membrane. ${\bf d}$, This effect can be used to dynamically transfer energy thus information to the cavity membrane from the PnC WG. Only when the cavity membrane is excited at the difference frequency, energy is transferred to it (solid line in the lower panel). In the absence of this excitation, no energy transfer takes place (dashed line in the lower panel).}
\label{fig 4}
\vspace{-1.1cm}
\end{center}
\end{figure*}

\setcounter{section}{0} %set all counters to 0, the first item is then Fig. 1 etc...
\setcounter{figure}{0}
\setcounter{equation}{0}
\renewcommand{\thesection}{S\arabic{section}}
\renewcommand{\thesubsection}{S\arabic{section}.\arabic{subsection}}
\renewcommand{\figurename}{\small{Figure S}\hspace{-0.11cm}}
\renewcommand{\theequation}{S\arabic{equation}}
\onecolumngrid %switch to single-column layout
\clearpage %leave the rest of the page blank and start a new one

\begin{center}

\noindent \textbf{{\Large Supplementary Information}}\\

\vspace{0.3cm}

\noindent{{\Large Phononic crystal waveguides for electromechanical circuits}}\\

\vspace{0.3cm}

\noindent \small{{\large D. Hatanaka, I. Mahboob, K. Onomitsu and H. Yamaguchi}}\\

\vspace{0.3cm}

\noindent \small{\large\it NTT Basic Research Laboratories, NTT Corporation, Atsugi-shi, Kanagawa 243-0198, Japan}\\

\end{center}

\vspace{0.2cm}

\section{Theoretical group velocity}
\hspace*{1em}In this section, a theoretical treatment is developed based on the formalism described in ref. 24 in order to calculate the group velocities in the PnC WG. The PnC WG consists of a 1-dimensional array of coupled membrane resonators that can be approximated by a chain of flexural mechanical beam resonators where each resonator unit is coupled to its neighbours via a mechanical spring as shown in Fig. S1a. The corresponding dispersion relation for this chain of articulated beams can be described as
\begin{equation}
\operatorname{Re}[kl]=\operatorname{Im}[\ln{\frac{D+\sqrt{D^{2}-16}}{4}}],
\end{equation}
where ${\it k}$ and ${\it l}$ are the wavevector and unit length of the beam and ${\it D}$ is given by
\begin{equation}
D=-a-\sqrt{a^{2}-4b+8}.
\end{equation}
In equation S2, ${\it a}$ and ${\it b}$ are given by
\begin{eqnarray}
\begin{split}
a=-2(\cosh \sqrt{\Omega}+\cos \sqrt{\Omega})-\frac{1}{2}\chi\sqrt{\Omega}(\sinh \sqrt{\Omega}-\sin \sqrt{\Omega}), \\
b=2+4\cosh \sqrt{\Omega}\cos \sqrt{\Omega}-\chi\sqrt{\Omega}(\cosh \sqrt{\Omega}\sin \sqrt{\Omega}-\sinh \sqrt{\Omega}\cos \sqrt{\Omega}),
\end{split}
\end{eqnarray}
with $\Omega$ and $\chi$ defined as
\begin{equation}
\Omega=\sqrt{\frac{\rho}{EJ}}l^{2}\omega_{\rm m},
\end{equation}
and
\begin{equation}
\chi=\frac{EJ}{Kl},
\end{equation}
where $\rho$, ${\it E}$, ${\it J}$ and $\omega$$_{\rm m}$ are the density, the Young's modulus, moment of inertia and angular frequency of the beam respectively and ${\it K}$ is the rotational spring constant. However, $\omega$$_{\rm m}$ is modified by the boundary conditions of the PnC WG to
\begin{equation}
\omega_{\rm m}=\sqrt{\omega^{2}-\omega_{\rm off}^2},
\end{equation}
since phonons are only allowed to propagate along the array but not in the direction perpendicular to it. This imposes a lower bound on the frequencies available where ${\it \omega}$ and ${\it \omega}$$_{\rm off}$ are the excitation and the cut-off frequencies respectively. The dispersion relation for the PnC WG can be determined by substituting equation S6 into equation S4, the result of which is plotted in Fig. S1b.\\
\hspace*{1em}The group velocity ${\it v}$$_{\rm g}$ in the PnC WG can then be readily determined from the identity
\begin{equation}
v_{g}=\frac{\partial \omega}{\partial k}
\end{equation}
using the dispersion relation plotted in Fig. S1b which is shown in Fig. S1c.
\begin{figure*}[floatfix]
\begin{center}
\vspace{-0.3cm}\hspace{8cm}
\includegraphics[scale=0.8]{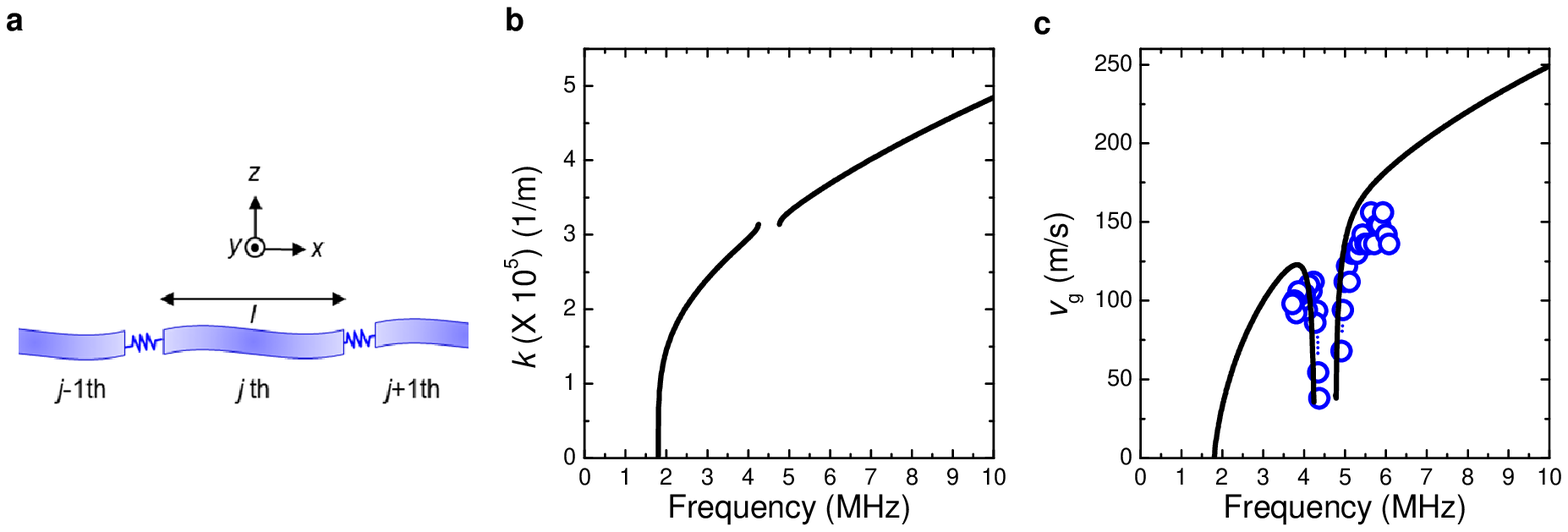}
\vspace{-17.5cm}
\caption{\textbf{The theoretical group velocity in the PnC WG.} ${\bf a}$, A schematic of the articulated array of beams used to extract the dispersion relation given in equation S1. Each unit consists of a flexural mechanical beam of length ${\it l}$ that is coupled to its neighbours via a mechanical spring. ${\bf b}$ and ${\bf c}$, The dispersion relation and the group velocity determined from the model described in ${\bf a}$ and equation S1 as a function of excitation frequency with $\rho$ = 5.36$\times$10$^{3}$ kg/m$^{3}$, ${\it E}$ = 8.53 $\times$ 10$^{10}$ Pa, ${\it l}$ = 1.00 $\times$ 10$^{-5}$ m, ${\it J}$ = 5.31 $\times$ 10$^{-15}$ kg$\cdot$m$^{2}$, $\chi$ = 0.160 and $\omega$$_{\rm off}$/2$\pi$ = 1.80 $\times$ 10$^{6}$ Hz. Open circles are the experimental group velocities in the PnC WG with ${\it N}$ = 100 as described in Fig. 3.}
\label{fig S1}
\vspace{-0.5cm}
\end{center}
\end{figure*}
\vspace{0.5cm}
\begin{figure*}[floatfix]
\begin{center}
\vspace{-1cm}\hspace{8cm}
\includegraphics[scale=0.8]{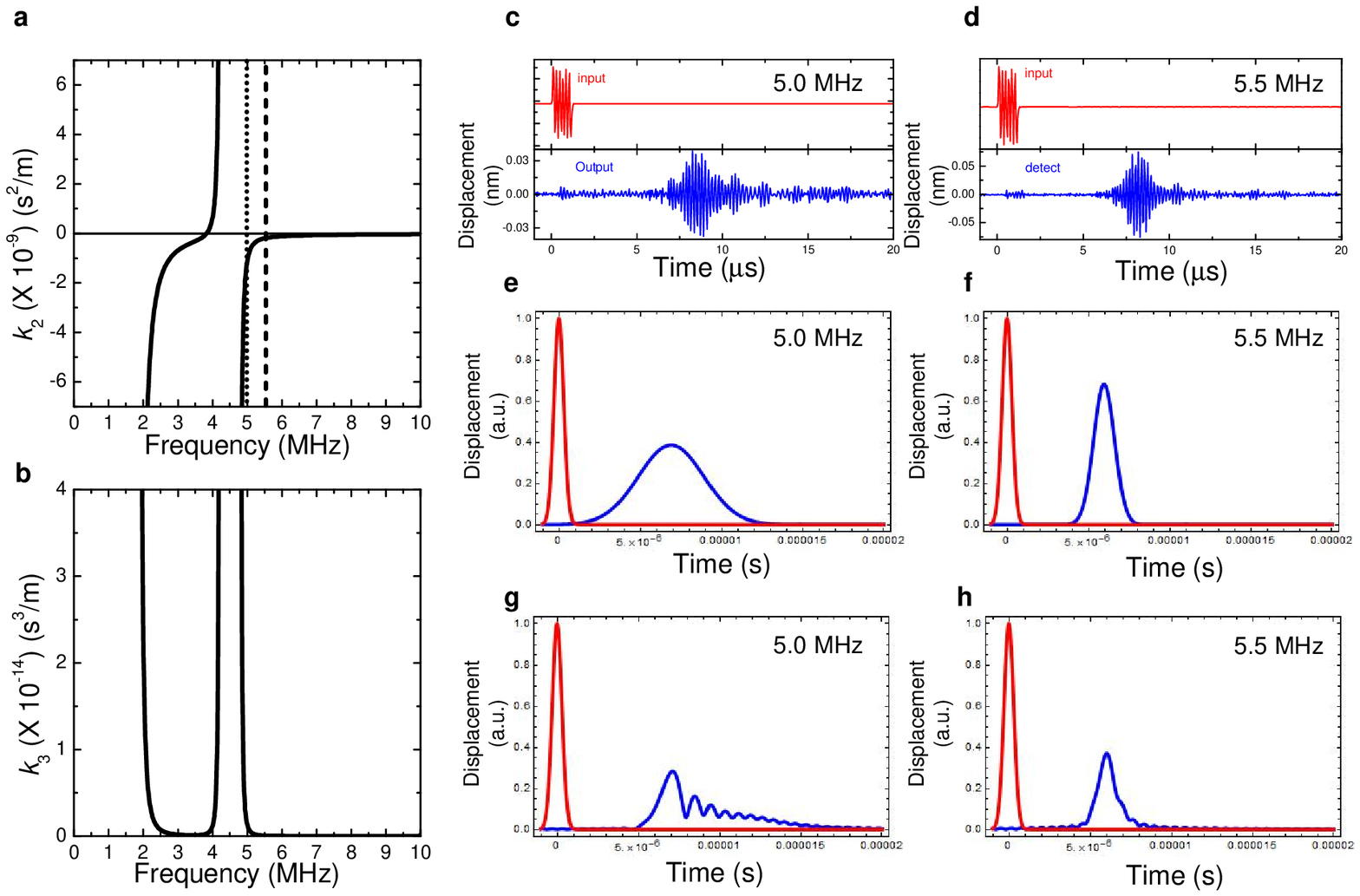}
\vspace{-12cm}
\caption{\textbf{Propagation dynamics of mechanical oscillations in a dispersive elastic media.} ${\bf a}$ and ${\bf b}$, The 2nd-order (${\it k}$$_{2}$) and 3rd-order (${\it k}$$_{3}$) dispersion relations respectively in the PnC WG as function of excitation frequency. ${\bf c}$ and ${\bf d}$, The experimental response of the mechanical oscillations at 5.0 and 5.5 MHz at the left edge in the PnC WG with ${\it N}$ = 100 respectively when excited at the right edge with a 1 $\mu$s pulse. ${\bf e}$ and ${\bf f}$ (${\bf g}$ and ${\bf h}$), The calculated response of the mechanical oscillation envelope from equation S16 (S17) at 5.0 and 5.5 MHz is shown in the blue lines with ${\it k}$$_{\rm 2}$ $\neq$ 0 and ${\it k}$$_{\rm 3}$ = 0 (${\it k}$$_{3}$ $\neq$ 0). A Gaussian pulse with a width of 1 $\mu$s is used as the input and is shown in the red lines.}
\label{extended data figure 2}
\vspace{0cm}
\end{center}
\end{figure*}
\section{Phonon pulse-shape evolution in the PnC WG}
\hspace*{1em}In order to develop an understanding of the details of the phonon propagation in the PnC WG, a model is developed in analogy to ref. S1 which describes the evolution of travelling mechanical oscillations in dispersive elastic media. Starting from the arbitrary dispersion function, $k(\omega)$,
the general form of the wave function is described by using the Fourier transform $\tilde{z}(\omega)$ as
\begin{equation}
z(x,t)=\frac{1}{2\pi}\int_{-\infty}^{\infty}\tilde{z}(\omega)\exp[i\{k(\omega)x-\omega t\}]d\omega.
\label{equ:S8}
\end{equation}
We consider the case that the propagating wave packet consists only of the Fourier components near the centre frequency $\omega_0$. Then  $k(\omega)$ can be expanded in the vicinity of $\omega_0$ as
\begin{equation}
k(\omega)=k_0+k_{1}{\Delta\omega}+\frac{1}{2}k_{2}{\Delta\omega}^{2}+\frac{1}{6}k_{3}{\Delta\omega}^{3}+\cdots \equiv k_0+k'(\Delta\omega),
\label{equ:S9}
\end{equation}
where $\Delta\omega=\omega-\omega_0$. The expansion coefficients $k_m$ are obtained by
\begin{equation}
k_{m}=\left(\frac{\partial^{m} k(\omega)}{\partial \omega^{m}}\right)_{\omega=\omega_{0}}    (m=0, 1, 2, 3, \cdots).
\end{equation}
By substituting \ref{equ:S9} into \ref{equ:S8}, we obtain
\begin{equation}
z(x, t)=z_{1}(x, t)\exp[i(k_{0}x-\omega_{0}t)]
\end{equation}
with
\begin{equation}
z_{1}(x, t)=\frac{1}{2\pi}\int_{-\infty}^{\infty}\tilde{z}_0 (\Delta\omega)\exp[i\{k'(\Delta\omega)x-{\Delta\omega}t\}]d\Delta\omega,
\label{equ:S12}
\end{equation}
where $\tilde{z}_0 (\Delta\omega)\equiv\tilde{z} (\omega_0+\Delta\omega)$ is the input pulse and ${\it z}_{\rm 1}$(${\it x}$, ${\it t}$) determines the envelope of the propagating waves.\\
\hspace*{1em}The group velocity in the PnC WG also exhibits a nonlinear correlation with respect to frequency (see Fig. 3 and Fig. S1c) and it arises from the non-negligible contribution of higher-order dispersions. In order to model the effect of higher-order dispersion on the pulse shape evolution, the contribution of 2nd-order dispersion is first considered. In equation \ref{equ:S9}, ${\it k}$$_{3}$ is neglected and the time ${\it t}$ is scaled as ${\it T}$ = ${\it t}$ - ${\it k}_{1}{\it x}$ which modulates equation \ref{equ:S12} to
\begin{equation}
z_{1}(x, T)=\frac{1}{2\pi}\int_{-\infty}^{\infty} \tilde{z}_{0}(\Delta\omega)\exp[i(\frac{1}{2}k_{2}\Delta\omega^{2}x-{\Delta\omega}T)]d\Delta\omega.
\end{equation}
Here, a Gaussian input pulse is used for $\tilde{z}$$_{0}$($\Delta\omega$) given by
\begin{equation}
\tilde{z}_{0}(\Delta\omega)=\int_{-\infty}^{\infty} z_{0}(T)\exp(i{\Delta\omega}T)dT,
\end{equation}
where
\begin{figure*}[floatfix]
\begin{center}
\vspace{-1cm}\hspace{8cm}
\includegraphics[scale=0.8]{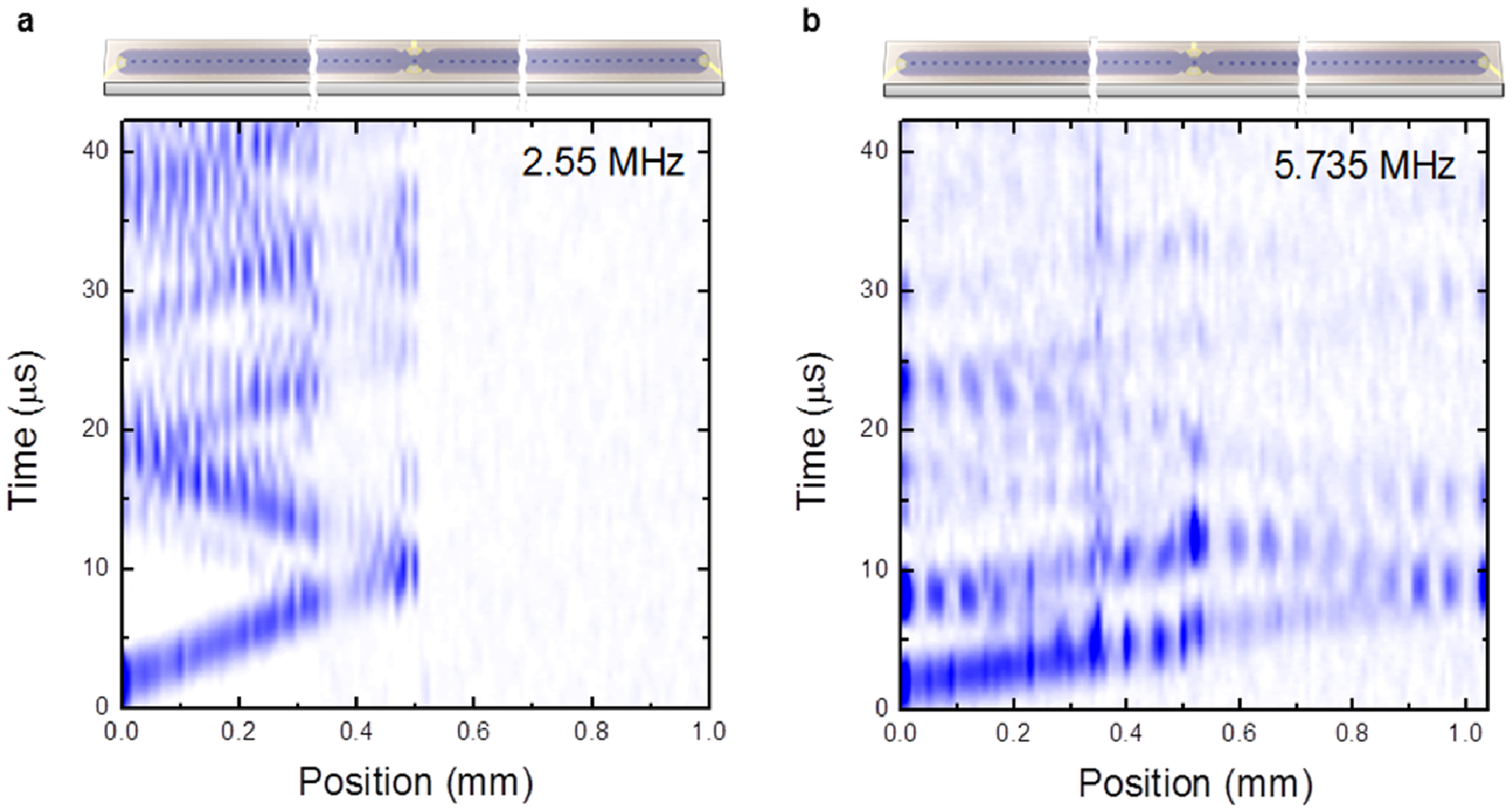}
\vspace{-14.5cm}
\caption{\textbf{Propagation dynamics of mechanical oscillations in the cavity-PnC WG} ${\bf a}$, The spatial and temporal response of the 2.55 MHz mechanical oscillation at various positions along the cavity-PnC WG. The pulse of the mechanical oscillation at 2.55 MHz is reflected by the cavity and is thus confined to the WG on the left side. ${\bf b}$, On the other hand, some of the mechanical oscillations at 5.735 MHz can propagate through the cavity.}
\label{extended data figure3}
\vspace{-0.5cm}
\end{center}
\end{figure*}
\begin{figure*}[floatfix]
\begin{center}
\vspace{-1cm}\hspace{20cm}
\includegraphics[scale=0.8]{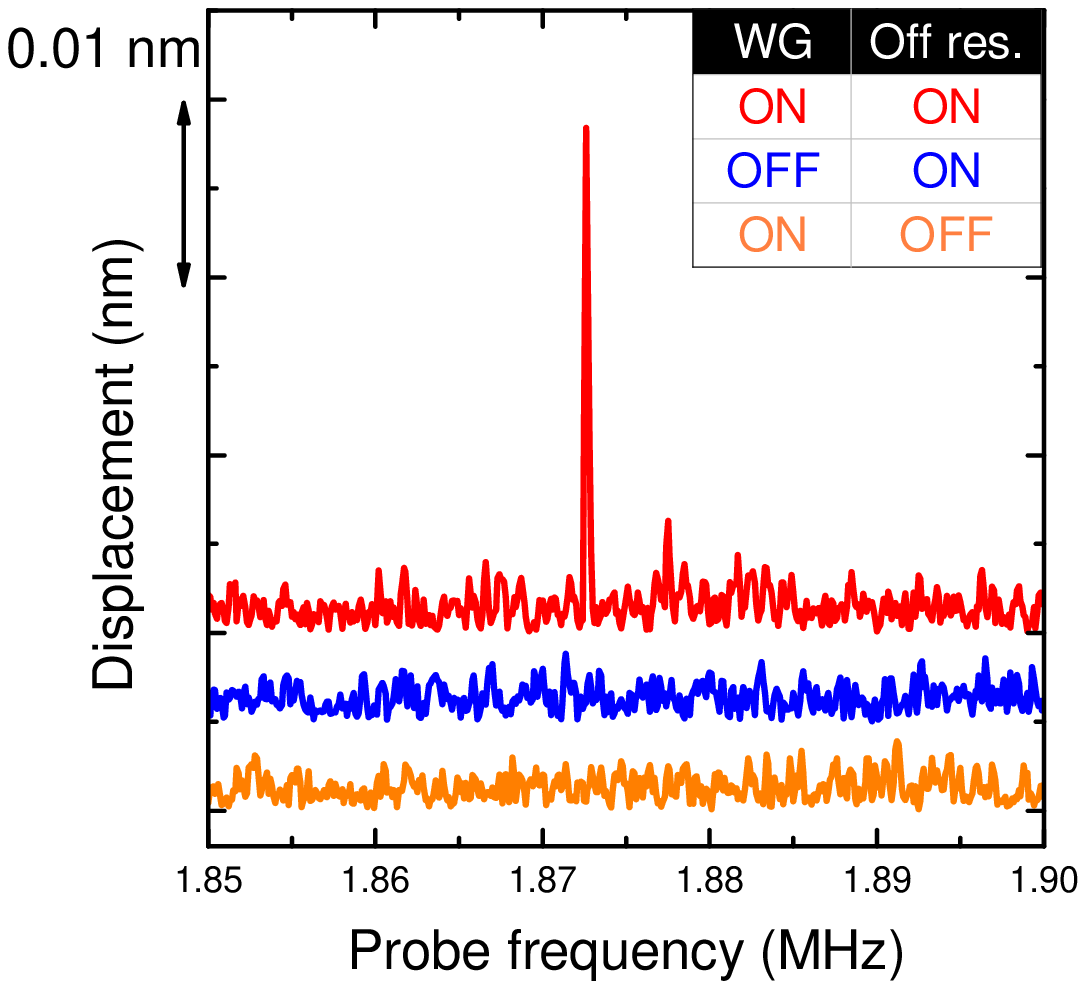}
\vspace{-13cm}
\caption{\textbf{The response of the cavity when dynamically coupled to the PnC WG.}The response of the localised cavity mode is measured as function of the WG mode at 5.745 MHz when the cavity is excited off resonance at 3.809 MHz namely the frequency difference to the WG. The localised cavity mode at 1.872 MHz is only observed when both the above inputs are simultaneously activated and it corresponds to energy being transferred from the WG to the cavity. All the spectra are offset for clarity.}
\label{extended data figure 4}
\vspace{0cm}
\end{center}
\end{figure*}
\begin{equation}
z_{0}(T)=\exp\left(-\frac{T^{2}}{2T_{0}^{2}}\right),
\end{equation}
and ${\it T}$$_{0}$ is the half width at 1/${\it e}$-intensity point. Finally from equation S13-S15, the amplitude of the pulse envelope at position ${\it x}$ can be expressed as
\begin{equation}
z_{1}(x, T)=\frac{T_{0}}{\sqrt{T_{0}^{2}-ik_{2}x}}\exp\left[-\frac{T^{2}}{2(T_{0}^{2}-ik_{2}x)}\right].
\end{equation}
The evolution of the pulse shape in the PnC WG can be calculated from equation S16  where ${\it k}$$_{2}$ is extracted from Fig. S2a and ${\it x}$ is set to be 1.0 mm namely the left edge of the PnC WG with ${\it N}$ = 100. The resulting calculation with these parameters reveals that the mechanical pulse at 5.5 MHz does not broaden, and is shown in Fig. S2f, as ${\it k}$$_{2}$ is small at this excitation frequency (see dashed line in Fig. S2a). On the other hand, the mechanical pulse is dramatically broader at 5.0 MHz, and is shown in Fig. S2e, where ${\it k}$$_{2}$ has a larger value (see dotted line in Fig. S2a). The experimental responses of the pulses at 5.0 and 5.5 MHz, shown in Figs. S2c and S2d, confirm the results of these calculations.\\
\hspace*{1em}In addition, the short input pulse in the experiment (1 $\mu$s) possesses a broadband of frequencies and as a result the parameter $\Delta\omega$$^{3}$ in equation S9 cannot be neglected. Consequently, equation S12 with a 3rd-order dispersion results in
\begin{equation}
z_{1}(x, T)=\frac{1}{2\pi}\int_{-\infty}^{\infty}\tilde{z}_{0}(\Delta\omega)\exp\left[i\left(\frac{1}{2}k_{2}\Delta\omega^{2}+\frac{1}{6}k_{3}\Delta\omega^{3}\right)x-i{\Delta\omega}T\right]d\Delta\omega.
\end{equation}
The pulse shape evolution determined from equation S17 with the parameters extracted from Fig. S2a and S2b are shown in Fig. S2g and S2h. This calculation indicates that the presence of ${\it k}$$_{3}$ induces an additional oscillatory structure on the pulse envelope at 5.0 MHz that develops as the pulse travels further and is shown in Fig. S2g. This observation is confirmed in the experimental measurement shown in Fig. S2c. Consequently, a short pulse in a nonlinear dispersive media can activate the 3rd-order dispersion which can distort the shape of the travelling mechanical oscillations. 

\vspace{0.5cm}

\section{Mechanical oscillations in the cavity-PnC WG}
\hspace*{1em}The introduction of a cavity into the PnC WG, as shown in Fig. S3a, creates a bottle neck between the WGs either side of the cavity. As a result, mechanical oscillations in the WG with frequencies $<$ 3 MHz experience a large acoustic impedance mismatch at the cavity interface and the resulting reflection confines these oscillations to the left-hand side PnC WG as shown in Fig. S3a. For mechanical oscillations with frequencies $>$ 3 MHz hence shorter wavelengths, the acoustic impedance mismatch is reduced enabling partial transmission through the cavity although significant reflection still takes place at the cavity interface as shown in Fig. S3b. These observations suggest that modifying the dimensions of the cavity membrane should enable the transmission properties of the PnC WG to be controllably engineered.\\

\section{Transparency in the cavity-PnC WG}
\hspace*{1em}An electromechanically-induced transparency is generated in the WG mode when the cavity is excited off-resonance as detailed in Fig. 4c$^{21,28,29}$. This off-resonance excitation of the cavity creates interference between the WG mode and the cavity localised mode which results in energy being transferred from the WG to the cavity.\\
\hspace*{1em}In order to confirm this, the spectral region of the localised mode of the cavity is probed as a function of the WG mode and the off-resonance cavity excitation as shown in Fig. S4. This measurement confirms that energy is transferred to the cavity thus activating its localised mode only when both inputs are activated. If one of the inputs is not active, energy cannot be transferred from the WG to the cavity.
\vspace{0.75cm}

\noindent \textbf{\Large{Reference}}

\vspace{0.2cm}

\noindent $^{S1}$ Agrawal, C. P. {\it Nonlinear Fiber Optics} (Academic Press, San Diego, 2001).\\

\end{document}